\documentclass[12pt,a4paper]{article}
\usepackage{latexsym,graphicx}
\newcommand{\eqref}[1]{(\ref{#1})}

\newcommand{\be}{\begin{equation}}
\newcommand{\ee}{\end{equation}}
\newcommand{\bea}{\begin{eqnarray}}
\newcommand{\eea}{\end{eqnarray}}

\begin{document}
\title{On the evolution of an entangled lepton-neutrino pair }
\author{B. Mesz\'ena and A. Patk\'os\\
Institute of Physics, E\"otv\"os University\\
H-1117, P\'azm\'any P\'eter s\'et\'any 1/A, Budapest, Hungary}
\vfill
\maketitle
\begin{abstract}
The evolution of the entangled muon-neutrino system emerging from charged
 pion decay is explored both in vacuum and in matter. The study is based on 
a Weisskopf-Wigner type wave-packet description. Explicit formulae are derived displaying modulation and 
attenuation of the oscillations
due to additional time scales characterising the production process. 
The case of neutrinos disentangled due to the detection of the muon 
 is also considered.
\end{abstract}
\section{Introduction}

The analysis of the neutrino oscillation phenomena necessarily includes beyond the propagation, also the production and detection stages, 
since the finiteness of the space-time
distance between the starting and the final events  is an essential part of the phenomenon. 
This feature leads to the smearing of the energy and the momentum of the particles and requires a wave packet 
description. 
Further time scales characterising the production and the detection 
appear, which modulate and damp the fundamental oscillations. These aspects
are constantly discussed in the literature of the past decades\cite{kaiser81,lipkin95,nauenberg99,akhmedov09,akhmedov10}. 
Terrestrial observability of decoherence 
effects was estimated very recently by Kayser and Kopp\cite{kaiser10}. This is the latest development in the ongoing attempts to extract more information from the oscillation experiments beyond the oscillation length \cite{lisi00,blennow05}.

Recent investigations focus on the entanglement aspects of the lepton-neutrino pair propagation in different "experimental" situations\cite{cohen09}.
Most recently Wu {\it et al.}\cite{boyan10} analysed the dependence of the density matrix on 
the detection time of the accompanying lepton when it is detected in addition to the propagation 
time of the neutrino. Their analysis of a two-body decay producing an entangled lepton-neutrino pair relies on first order perturbation 
theory which is strictly valid only for times shorter than the lifetime of the mother-particle.  This approach seems to be more relevant for long-lived
sources (for example for beta decay of long-lived radiactive isotopes). Decohering 
effects reflecting the finite width of a Gaussian wave packet were analyzed in Ref.\cite{giunti98}. Decoherence resulting from
 the finite lifetime of the mother-particle was systematically investigated in 
Refs.\cite{grimus96,grimus98}. Both studies considered disentangled neutrinos.

Our goal in this paper is to derive simple expressions governing the coherence of the system in a unified 
treatment which includes the effects of all relevant
timescales and relies on an approach which makes use of wave packets. In our discussion we make use of
 two-body decays and the description  is valid for times much larger than the source's lifetime. An important process of this type is the decay of charged pions
$\pi\rightarrow \mu + \nu_\mu$, to which we shall mostly refer in this Letter.
The treatment will be extended also to the propagation through matter where density dependent resonant effects 
increase of the coherence length is observed, which reaches even infinity for some specific value of the lepton density. 
Consequences of muon detection will be carefully explored and compared
to the oscillations displayed by an entangled neutrino.

The Letter starts by an overview of the results of the
entangled wave-packet treatment of the 2-flavor muon+neutrino 
propagation following Ref.\cite{nauenberg99}. 
Simple approximate formulae which describe 
the modulation and decoherence of the basic oscillations both in vacuum and matter represent the main results of our note. In the second part
we repeat the analysis for the case, when the muon is detected near the decay tube of the pion. 
We achieve also in this case a unified picture of the effects of finite lifetime and the period of the oscillations observed due to the 
finite propagation time.

\section{Wave-packet evolution without observing the entangled muon}

The state vector of an entangled collinearly outflying muon-neutrino pair
in the Weisskopf-Wigner approximation reads as follows \cite{nauenberg99}:
\begin{eqnarray}
|\Psi(x_\nu,x_\mu,t)\rangle=(\psi_1(x_\nu,x_\mu,t)\cos\Theta|m_1\rangle
+\psi_2(x_\nu,x_\mu,t)\sin\Theta|m_2\rangle)|\mu\rangle,&&\nonumber\\
\psi_i(x_\nu,x_\mu,t)={\cal N}\int dp_\nu\int dp_\mu f_\pi(p_\nu+p_\mu)
e^{i(p_\nu x_\nu+p_\mu x_\mu)-i(E_{\nu_i}+E_\mu)t}
\frac{1}{E_{\nu_i}+E_\mu-E_\pi+iM_\pi\Gamma/2E_\pi},&&
\label{state-vector}
\end{eqnarray}
where $E_{\nu_i}^2=m_i^2+p_\nu^2, E_\pi^2=M_\pi^2+(p_\nu+p_\mu)^2$ and the state vector of the muon-type neutrino created from 
the decay is $|\nu_\mu\rangle =\cos\Theta|m_1\rangle 
+\sin\Theta|m_2\rangle$. The label $m_i$ refers to the mass of the
 i-th mass eigenstate, $f_\pi(p)$ is the momentum profile, $\Gamma$
 is the energy width of the
decaying pion, which will be later
 assumed to have zero momentum on the average. 
${\cal N}$ is an appropriate normalisation constant which ensures that
$\int dx_\nu\int dx_\mu |\psi_i|^2=1$ is fulfilled.  Eq.(\ref{state-vector}) arises from the complete
Weisskopf-Wigner expression when $t>>\Gamma^{-1}$. (It should be contrasted with the treatment of Ref.\cite{boyan10}
which is valid oly for early times $t<<\Gamma^{-1}$.)

The survival probability of $\nu_\mu$ can be studied by projecting 
(\ref{state-vector}) on the initial (muon + muon-neutrino) state. The spatial propagation of the $\mu$-type neutrino profile is
 tracked when the muon is not detected, e.g. one integrates the absolute square of the projection over $x_\mu$:
\begin{equation}
Prob(x_\nu,t)=\int dx_\mu\left(\cos^4\Theta|\psi_1|^2+
\sin^4\Theta|\psi_2|^2+\frac{1}{2}\sin^2(2\Theta){\textrm Re}
(\psi_1^*\psi_2)\right).
\label{prob-nu}
\end{equation}
Even simpler is the question about the survival of $\nu_\mu$ without its localisation, e.g. integrating also over $x_\nu$.
 The oscillating part of this probability 
is proportional to the real part of the following integral:
\begin{eqnarray}
I_{12}=\int dx_\nu\int dx_\mu\psi_1^*\psi_2
&=&\int dp_\nu\int dp_\mu
|f(p_\nu+p_\mu)|^2e^{i(E_{\nu_1}-E_{\nu_2})t}\nonumber\\
&\times&\frac{1}
{(E_{\nu_1}+E_\mu-E_\pi+iM_\pi\Gamma/2E_\pi)
(E_{\nu_2}+E_\mu-E_\pi-iM_\pi\Gamma/2E_\pi)}.
\label{Eq:i12}
\end{eqnarray}
Another way to recognize its interest is to realise that it
 determines the off-diagonal element of the
flavor density matrix in the mass eigenbasis, and characterizes this way
the coherence of the two-state system:
\begin{equation}
\rho^{(m)}_{12}(t)=\frac{1}{2}I_{12}\sin(2\Theta).
\end{equation}
Oscillation phenomena actually are governed by this component. 
The experimentally observed oscillation in the diagonal matrix elements in the flavor or interaction eigenbase ($\rho^{(f)}$) is 
just the consequence of the orthogonal transformation connecting the two bases.

A simple and numerically accurate approximate representation of $I_{12}$
can be derived following \cite{nauenberg99} by expanding all factors of the integrand of (\ref{Eq:i12}) 
to linear order around some sort of "mean" momenta (denoted by capital letters): $P_\nu,
P_\mu$, and $P_\pi$. They are determined from the requirement of energy and momentum conservation 
upon neglecting the mass of the neutrino. After the expansion one demonstrates with an appropriate change of 
variables that the two-variable integral factorizes and one integration even can be performed 
via Cauchy's theorem. The final result is of the following form:
\begin{eqnarray}
&\displaystyle
I_{12}=\exp\left(i\frac{\Delta m^2}{2P_\nu}t\right)
\frac{2\pi}{1-i\frac{\Delta m^2}{2P_\nu}\frac{E_\pi}{M_\pi\Gamma}}
F(t)\exp\left(-\frac{\Delta m^2}{4P_\nu^2}\frac{M_\pi\Gamma}{E_{\pi}(1-v_\mu)}t\right),
\nonumber\\
&\displaystyle
F(t)=\int dp|f(p)|^2\exp\left(-i\frac{\Delta m^2}{2P_\nu^2}\frac{v-v_\mu}
{1-v_\mu}(p-P_\pi)t\right).
\label{factor-form1}
\end{eqnarray}
Here $\Delta m^2$ is the squared mass difference of the two neutrinos, and 
$v=P_\pi/E_\pi, v_\mu=P_\mu/(P_\mu^2+m_\mu^2)^{1/2}$.   
The second factor in this expression suppresses the amplitude of the oscillations only when the lifetime of 
the mother-particle is much larger than the oscillation time, which is not the case of the pion decay. Clearly, the integral 
$F(t)$ represents the
Fourier transform of the wave function profile of the pion, and implies the decay of the oscillating term with characteristic size $1/d$.
The last factor provides attenuation of the oscillation with some 300 years characteristic decay time when one makes use of realistic parameters \cite{grimus98}. 
The coherence length emerging from the concurrence of the last two factors is determined by $max (\Gamma, d)$. 
Below we shall assume $P_\pi=0$, since then the formulae look much simpler.

Modified formulae reflecting medium effects can be written in an analogous form. Starting 
with a Hamilton operator which contains the Fermi interaction of the e-type neutrinos with 
the constituents of the medium\cite{sigl93} one solves the Schr\"odinger equation of the density matrix. 
 For the leading effect the mass difference in the denominator can be neglected 
(in both factors we write for the neutrino energy its value calculated at $P_\nu$). 
Then for the probability of the flavor flip one finds
\begin{eqnarray}
P(\nu_\mu\rightarrow \nu_e,t)&=&(2\pi N)^2\int dp_\nu\int dp_\mu
\left(\frac{\Delta m^2}{\Delta m_{eff}^2(p_\nu)}\right)^2|f(p_\nu+p_\mu)|^2
\nonumber\\
&\times&\frac{(1-\cos[(\Delta m_{eff}^2(p_\nu)/2p_\nu)t])\sin^2(2\Theta)/2}
{|E_{P_\nu}+E_\mu-E_\pi-i\Gamma/2|^2}.
\end{eqnarray}
Here we have introduced the medium and momentum dependent effective squared mass splitting
\begin{equation}
\Delta m_{eff}^2(p)=\sqrt{(\Delta m^2)^2+c^2p^2-2cp\Delta m^2\cos(2\Theta)},
\qquad c=2\sqrt{2}NG_F
\end{equation}
($G_F$ is the Fermi constant, $N$ the density of the medium).

\begin{figure}[!t]
\centerline{ 
\includegraphics[keepaspectratio,width=0.3\textwidth,angle=0]{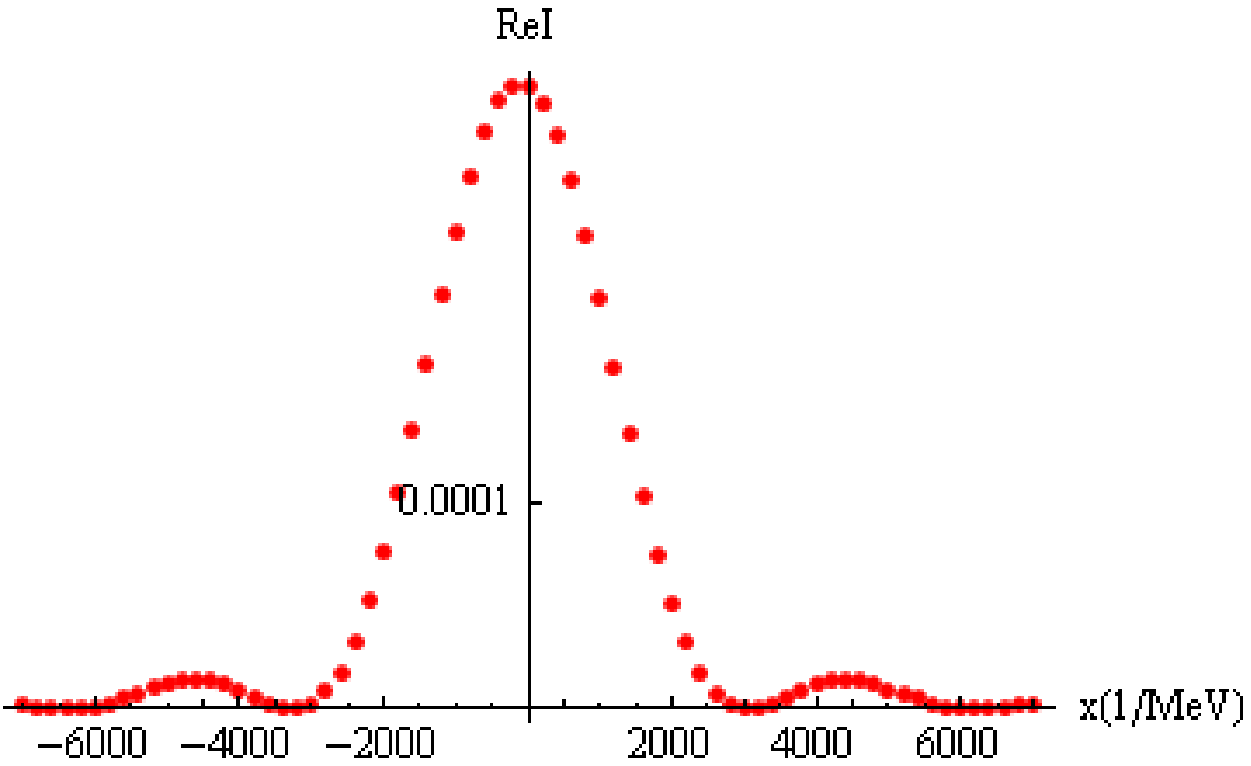}
\includegraphics[keepaspectratio,width=0.3\textwidth,angle=0]
{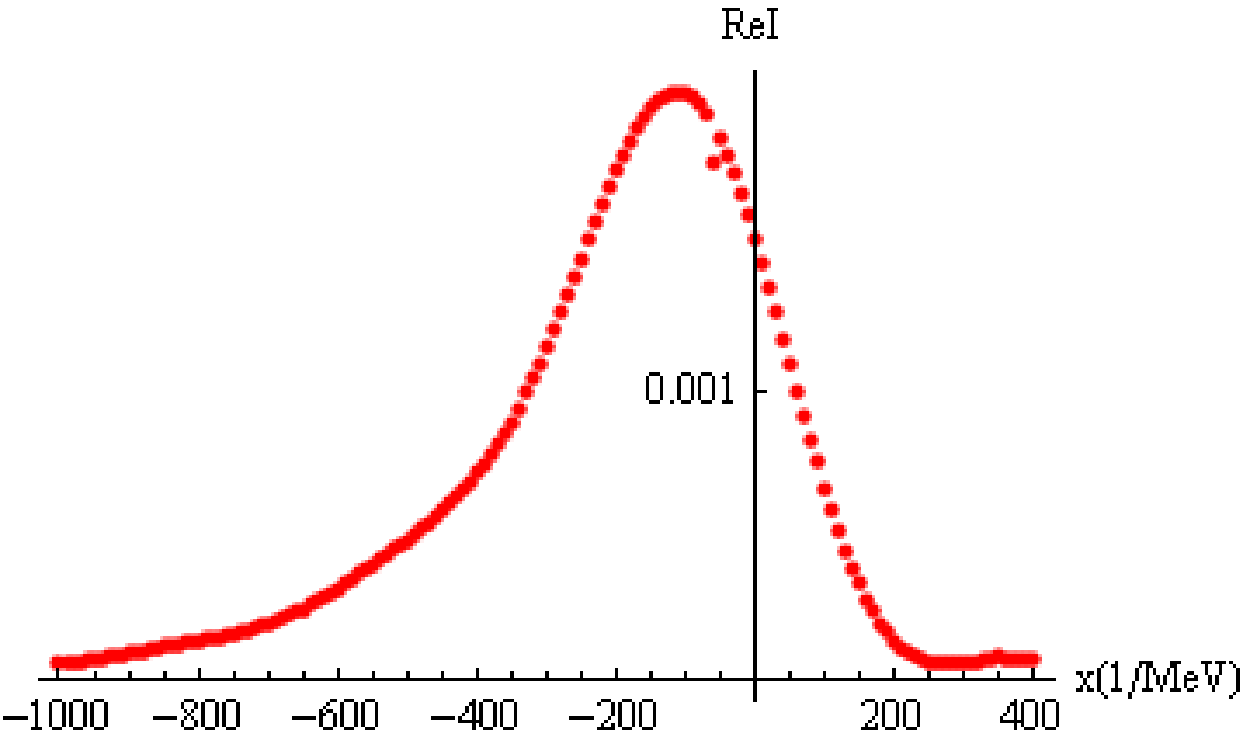}
\includegraphics[keepaspectratio,width=0.3\textwidth,angle=0]
{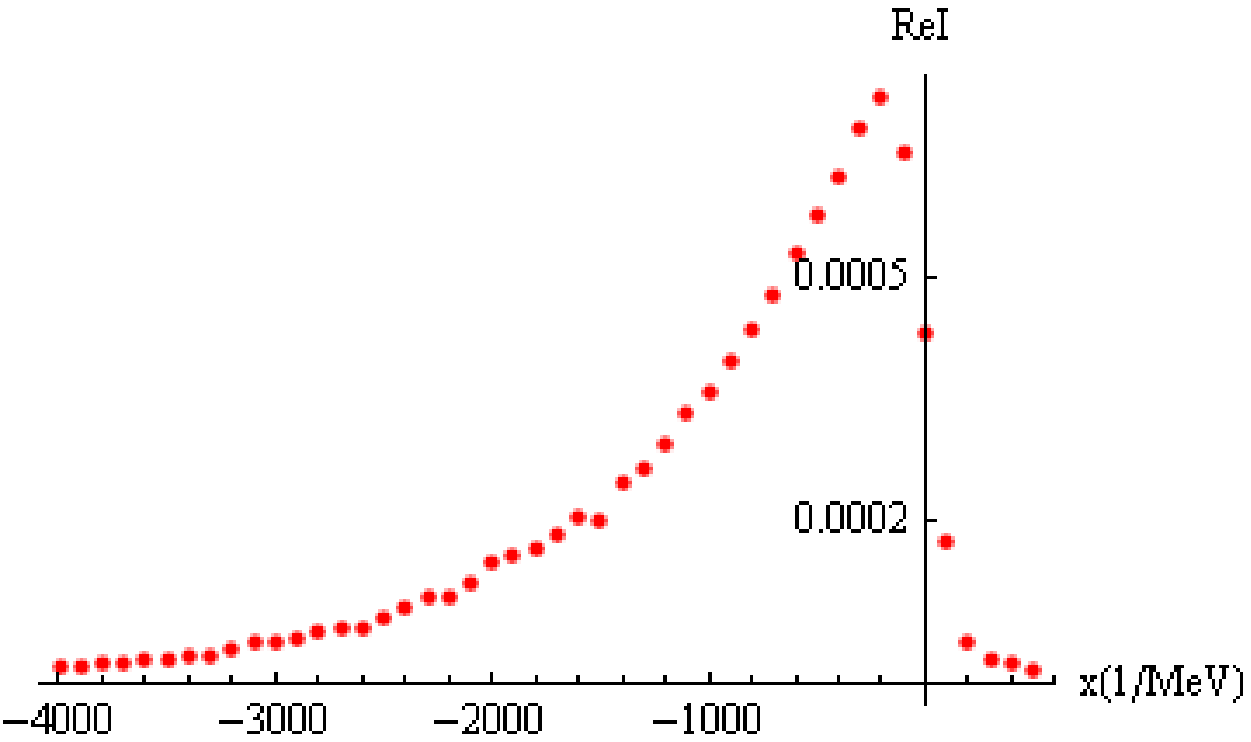}}
\caption{The starting ($t=0$) probability distribution of a rightward propagating 
neutrino wave-packet. The inverse lengths characteristic for the figures 
from the left to the right are the following (measured in MeV): $(\Gamma, d)=(0.01,0.001),
(0.005,0.01), (0.001,0.01)$, respectively.
}
\label{Fig:neutrino-signal}
\end{figure}

Assuming slow variation of the effective mass splitting over the momentum range of the neutrino, 
the double integral can be evaluated with the same approximate technique as explained for (\ref{factor-form1}). 
It leads to the following factorized form:
\begin{eqnarray}
&\displaystyle
P(\nu_\mu\rightarrow \nu_e,t)=\left(\frac{\Delta m^2}{2\Delta m_{eff}^2(P_\nu)}
\right)^2\sin^2(2\Theta)(1-{\textrm Re}I_{matter}),\nonumber\\
&\displaystyle
I_{matter}=2\pi
\exp\left(i\frac{\Delta m_{eff}^2(P_\nu)}{2P_\nu}t\right)F_{matter}(t)
\exp\left(-\frac{|\Delta
\tilde m^2|}{4P_\nu^2}\frac{\Gamma}{1-v_\mu}t\right).
\end{eqnarray} 
The absence of the second factor of the analogous expression (\ref{factor-form1}) tells that here we discuss 
the case when the lifetime of the pion is much shorter than the oscillation period.
$F_{matter}$ has the same form as in (\ref{factor-form1}) with a substitution $\Delta m^2\rightarrow \Delta\tilde m^2$, where
\begin{equation}
\Delta\tilde m^2=\frac{\Delta m^2(\Delta m^2-cP_\nu\cos(2\Theta))}
{\Delta m_{eff}^2(P_\nu)}.
\end{equation}
As a function of $c$ (e.g. the matter density) $\Delta\tilde m^2$ is nonmonotonic and with realistic parameters for $N\approx 2\times 10^2N_A/{\textrm cm}^3$ it reaches zero ($N_A$ is the Avogadro number). 
It is worthwhile to emphasize that the quantities influencing the variation of the oscillation time and of the coherence length in matter are different.

Although with the integration over $x_\nu$ information was lost on the
location of the neutrino, one might estimate it intuitively with the identification $\bar x_\nu\approx t$. More detailed information can be obtained by evaluating (\ref{prob-nu}), which we are going to discuss next. 

In order to ease the triple momentum integral in its last term occuring after the $x_\mu$ integration, we have assumed that $f(p)$ is just a window function of size $2d$. The result still reflects all generic features not depending on the specific form of the pion profile. In this specific case
 by careful mapping of the range of variation the $p_\mu$ integration can be performed analytically. Denoting its (not-quite-transparent) result by $G(p_\nu,q_\nu)$ 
one reduces the integral to the form
\be
I(x_\nu,t)=2\pi N^2\int\int_{|p_\nu-q_\nu|<2d}dp_\nu dq_\nu G(p_\nu,q_\nu)
\exp[i(q_\nu-p_\nu)x_\nu-i(E_{\nu 1}(q_\nu)-E_{\nu 2}(p_\nu))t].
\label{Eq:nu-wave}
\ee
The remaining two integrations were performed numerically. The shape of the profile of the 
muon-type neutrino signal is obtained by adding to this term also the diagonal contributions.
The resulting probability distribution depends on the relation of $\Gamma$ and $d$, as can be seen from Fig. \ref{Fig:neutrino-signal}. 
The shape propagates with oscillating amplitude and the forms displayed in the Figure will reappear periodically with some changes reflecting decoherence.

When $d<<\Gamma$ the profile is symmetric to some central point which can be identified with the actual position of the neutrino. When one goes over to the opposite situation
where $d>>\Gamma$, the time-profile becomes asymmetric. Its forward slope
becomes very steep of size $\sim 1/d$, while the size of the back-tail is determined by $\Gamma^{-1}$. 
Unfortunately the resolution of the neutrino detectors will not allow for a long time the detailed mapping of the neutrino signal. Still this part of our analysis shows that the detection rate might be influenced by the way the wave function of the detector particle overlaps with the neutrino profile. 

\section{Effects of muon detection on the neutrino propagation}

In  realistic experiments the detection of the accompanying muons
usually provides the estimate for the neutrino flux generated in the 
decay tube of the pions. This disentangling event has an impact on 
the observations of the neutrino oscillations which we shall discuss below.

Our crude picture of muon detection assumes that the detection consists of measuring the momentum of the muon by a spatially extended detector. 
Let its measured  value be $Q$ and the measurement happens at time $t_\mu$.
This disentangling event produces the following one-particle wave function for the neutrino:
\begin{eqnarray}
&\displaystyle
|\nu_\mu\rangle=\phi_1\cos\Theta|m_1\rangle +\phi_2\sin\Theta |m_2\rangle,\nonumber\\
&\displaystyle
\phi_i(x_\nu,t)=n_i(Q)N\int dp_\nu \frac{f(p_\nu+Q)}{E_{\nu,i}+E_\mu(Q)-E_\pi+i\Gamma/2}
e^{i(p_\nu x_\nu-E_{\nu,i}t)} e^{iE_\mu(Q)t_\mu}.
\end{eqnarray}
Here $n_i(Q)$ is an appropriate normalisation factor for the one-particle wave function of the neutrino.
The exponent of the last factor is constant, therefore can be omitted. 

We wish to compare the oscillating part of the survival probability in
 this case to (\ref{Eq:i12}), therefore we integrate over $x_\nu$:
\begin{eqnarray}
I_{12}^{dis}(t)&=&2\pi n_1(Q)n_2(Q)N^2\nonumber\\
&\times&\int dp_\nu \frac{|f(p_\nu+Q)|^2e^{i(E_{\nu 1}-E_{\nu 2})t}}{(E_{\nu 1}+E_\mu(Q)-E_\pi-i\Gamma/2)
(E_{\nu 2}+E_\mu(Q)-E_\pi+i\Gamma/2)}.
\end{eqnarray}
This integral can be estimated analytically in two limiting cases: 
$d<<\Gamma$ and $d>>\Gamma$.
When $d<<\Gamma$, the momentum uncertainty is smaller therefore $p_\nu$ varies around $P_\nu=-Q$.
 The variation of the numerator is faster, therefore the denominator can be taken out as a constant from the integration. 
The requirement of the normalisation of $I_{12}^{dis}$ to unity when $\Delta m^2=0$ leads to the cancellation of these factors with
$n_1n_2$.
In the opposite limiting case $d>>\Gamma$ the uncertainty of the energy is small therefore the "mean" value of $p_\nu$ 
is calculated from the energy conservation:
$P_\nu=M_\pi-\sqrt{m_\nu^2+Q^2}$.  
The remaining integral is performed via Cauchy's theorem. 
The results of the integration in the two limiting cases are as follows:
\begin{eqnarray}
I_{12}^{dis}(d<<\Gamma)&\approx&2\pi\exp\left(-i\frac{\Delta m^2}{2Q}t\right)\int dp|f(p)|^2\exp\left(-i\frac{\Delta m^2}{2Q^2}pt\right),\nonumber\\
 I_{12}^{dis}(d>>\Gamma)&\approx&\exp\left(-i\frac{\Delta m^2}{2Q}t\right)\frac{1}{1+i\frac{\Delta m^2}{2Q\Gamma}}\exp
\left(-\frac{\Delta m^2\Gamma}{4Q^2}t\right).
\label{i12-disentangle}
\end{eqnarray}
It follows that the coherence length in case of the disentangled neutrino is determined by $min (\Gamma, d)$. One can easily find that physically 
this corresponds to the case when the coherence
is suppressed by the vanishing overlap of the different mass-components of the neutrino wave function. Its characteristic
time is determined by the difference of the group velocities of the two components.

In conclusion we summarize the findings of our note:
\begin{itemize}
\item
In case of the entangled muon+neutrino pair if $\Gamma$ and $d$ are of different order of magnitude then the coherence time/length is always shorter than 
the separation time of the two eigen-mass modes. In the disentangled case, however, this latter determines the coherence of the oscillations. 
\item
The dependence of the coherence length on the muon momentum and velocity is slightly different in the entangled and the disentangled cases (cf. (\ref{factor-form1}) and
(\ref{i12-disentangle})).
\end{itemize}
In an experimental situation only some fraction of the muons is detected because of the finite efficiency of the muon-detector. Therefore the neutrino beam realistically is a mixture of entangled and disentangled neutrinos, which will necessitate some refinement of the analysis when the experiments become sensitive to such details.
The measurement of the coherence length of the oscillations might extend the conventional oscillation paradigm in the not-too-distant future. Since
there is no universal characterisation of the decoherence, future measurements of the decoherence length 
might bring informations on the actual wave-packet profile and also on the degree of entanglement of the neutrino beam.

\section*{Acknowledgments}
This research was supported by the Hungarian Research Fund Grants K77534 and T068108.

\end{document}